\input{aipcheck}

\documentclass[
    ,final            
  ]
  {aipproc}

\layoutstyle{6x9}

\usepackage{rotating}

\begin{document}

\def\met{\mbox{\ensuremath{\, \slash\kern-.6emE_{\rm T}}}}

\title{Cosmic Ray Muons Timing in the ATLAS Detector}


\author{Bernhard Meirose \\ 
On behalf of the ATLAS Tile Calorimeter System
}{
  address={Albert-Ludwigs-Universit\"{a}t Freiburg \\
Hermann-Herder-Str. 3 79104 Freiburg i. Br., Germany}
}

\begin{abstract}
In this talk I discuss the use of calorimeter timing both for detector
commissioning and in searches for new physics. In particular I present 
real and simulated cosmic ray muons data (2007) results for the ATLAS Tile Calorimeter system.
The analysis shows that several detector errors such as imperfect calibrations can be uncovered. I also demonstrate the use of ATLAS Tile Calorimeter's excellent timing resolution in suppressing cosmic ray fake missing transverse energy ($\met{}$) in searches for supersymmetry.

\end{abstract}

\maketitle


\section{Introduction}

The ATLAS (A Toroidal LHC ApparatuS) experiment \cite{ATLAS}, one of the two LHC (Large Hadron Collider) \cite{LHC} general purpose detectors, is ready for data taking. Results from the LHC are expected to shed new light on the origin of electroweak symmetry breaking, and potentially open a wide spectrum of new physics discoveries. One of the most crucial steps in the commissioning of the ATLAS detector is the cosmic ray muon analysis. Before the LHC starts its operation by November 2009, cosmic rays remain the best source of real data physics analysis in ATLAS, testing the
stability of the detector and working out any bugs as early as possible. It also offers the possibility of developing techniques that will be used in real LHC data analysis.

\section{Tile Calorimeter Detector Unit}

The Tile Calorimeter \cite{Tile} is a sampling calorimeter with iron absorber and scintillating plastic ``tiles'' as
the active material, with a novel geometry of alternating layers perpendicular
to the beam direction. It is a cylindrical structure and is divided into three 
cylindrical sections; a barrel and two extended barrels. All the three sections are divided
in $\phi$ by 64 wedge modules. The light from each scintillating tile is transmitted on two sides by a
wavelength-shifting fibre, and the fibres are bundled together to form readout cells
with three different sampling depths. Each cell is read out by two photomultipliers (PMT's), one on either side, with a total of approximately 10000 PMT's in the entire TileCal. The main function of the Tile Calorimeter is to contribute to the  energy measurement of jets produced 
in proton-proton interactions and, with the addition of the end-cap and forward calorimeters, to provide a good measurement of missing transverse energy.

\section{Cosmic Muons Time-of-Flight and cell timing cuts}

As an example of ATLAS commissioning work with cosmics we calculate the time-of-flight (TOF) between two back-to-back (in $\phi$) Tile Calorimeter modules \cite{Meirose&Teuscher}. Typically muons deposit between 1 to 3 GeV in each projective tower of the Tile Calorimeter, compared to an electronics noise of O(100) MeV per tower. Knowing the distance between the cells (which are different for the three cell types A, BC and D) and that cosmic muons are close to the speed of light, we compared the TOF from real cosmic ray data recorded in the summer of 2007 with expected from geometry. Depending on their trajectory, cosmic ray muons traveling from the top to the bottom of ATLAS at near the speed of light will have a time-of-flight of typically 18-20 ns over their travel distance of approximately 6-7 m in the Tile Calorimeter. For real data, the time for each cell is the PMT time average, and the TOF is the time difference between a calorimeter top cell ($\phi$ > 0) and a bottom cell ($\phi$ < 0).  At the time of this calculation (2007), the PMT time inter-calibration for the Tile Calorimeter, which were obtained in LASER runs studies \cite{Clement}, were still at early stages. One important result of the work with cosmics was the discovery of small deviations in the TOF for some cells, which suggested some inter-time offsets could be improved. This is shown in Figure 1. Note that there is a slight asymmetry in the plot (towards -10 ns).  We stress the fact that this plot is a commissioning example and therefore, cannot be seen as a final result. All time inter-calibrations were corrected and currently one would get a narrower width around the expected mean of the distributions \cite{Fiorini}.

\begin{figure}
  \includegraphics[height=.25\textheight]{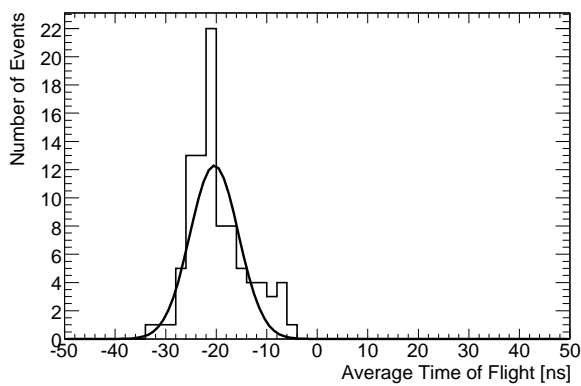}
  \caption{Time-of-Flight between D cells in back-to-back modules.}
\end{figure}

Another interesting application of calorimeter timing techniques is for signal-to-noise ratio improvement. For a real cosmic muon event both PMT's in a same Tile Calorimeter cell, register similar times and energies. This fact can be used to distinguish a low energy muon from electronics noise. Figure 2 shows the TOF distribution for A cells above 100 MeV (left) and the same distribution after a 6 ns PMT time-difference cut (right). Note that the peak around -16ns was poorly resolved before the timing cut. Although in this example the type of signal considered was cosmic-ray muons, Tile Calorimeter cell time-difference closeness is a characteristic of real physics events. This could be particularly important in LHC low energy events such as minimum-bias.

\begin{figure}
\begin{tabular}{c c}
  \includegraphics[height=.25\textheight]{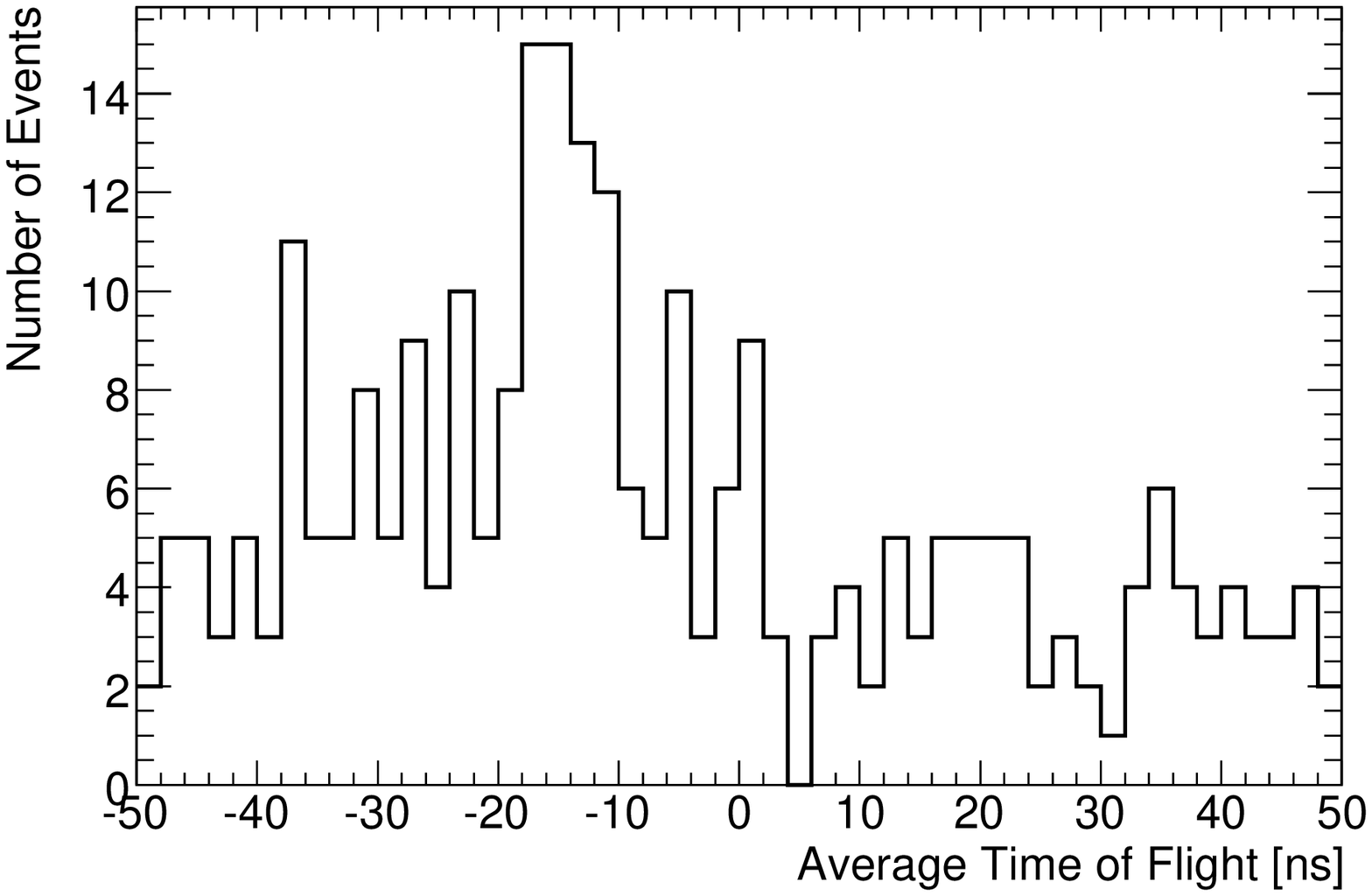}
&

  \includegraphics[height=.25\textheight]{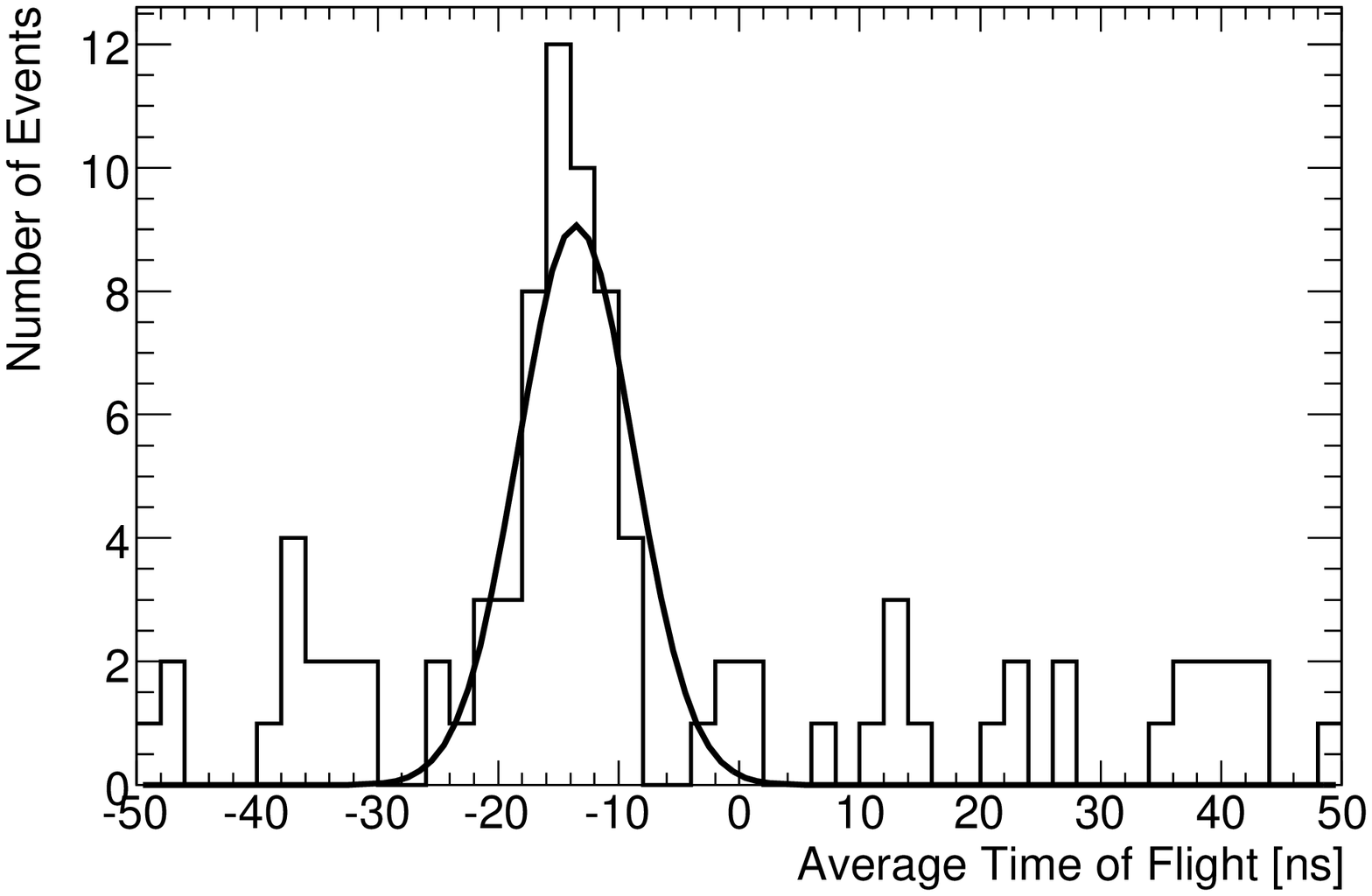}
\end{tabular}
  \caption{Time-of-Flight for Tile Calorimeter A cells with energy > 100 MeV before (left) and after (right ) a 6 ns PMT time difference cut.}

\end{figure}

\section{Fake $\met{}$ Rejection in SUSY searches}

One possible scenario for an early supersymmetry signature in ATLAS would be an excess of
events with large missing transverse energy. Therefore all sources that can fake  $\met{}$ represent
backgrounds to SUSY. Examples are: dead or noisy calorimeter channels, machine induced backgrounds
and cosmic rays, triggered by themselves or overlapping with QCD jets or minimum bias events. Because of its precision ($\sim$ 1 ns) Tile Calorimeter timing can be a powerful tool to remove fake  $\met{}$ \cite{CSC}. Cosmic ray events, are mostly single particles crossing the detector from top to bottom so, since cells on the top are hit first, the distribution of up-down TOF should peak at around -18 ns. However, for events coming from proton-proton collisions, the up-down time is a difference of arrival times of two different particles created at the same time, so that distribution averages at zero allowing for a clear discrimination between signal and background. Figure 3 shows the TOF for a simulated Monte Carlo sample of cosmic ray muons (left) and QCD dijet events (right).


\begin{figure}
\begin{tabular}{c c}
  \includegraphics[height=.3\textheight,angle=270]{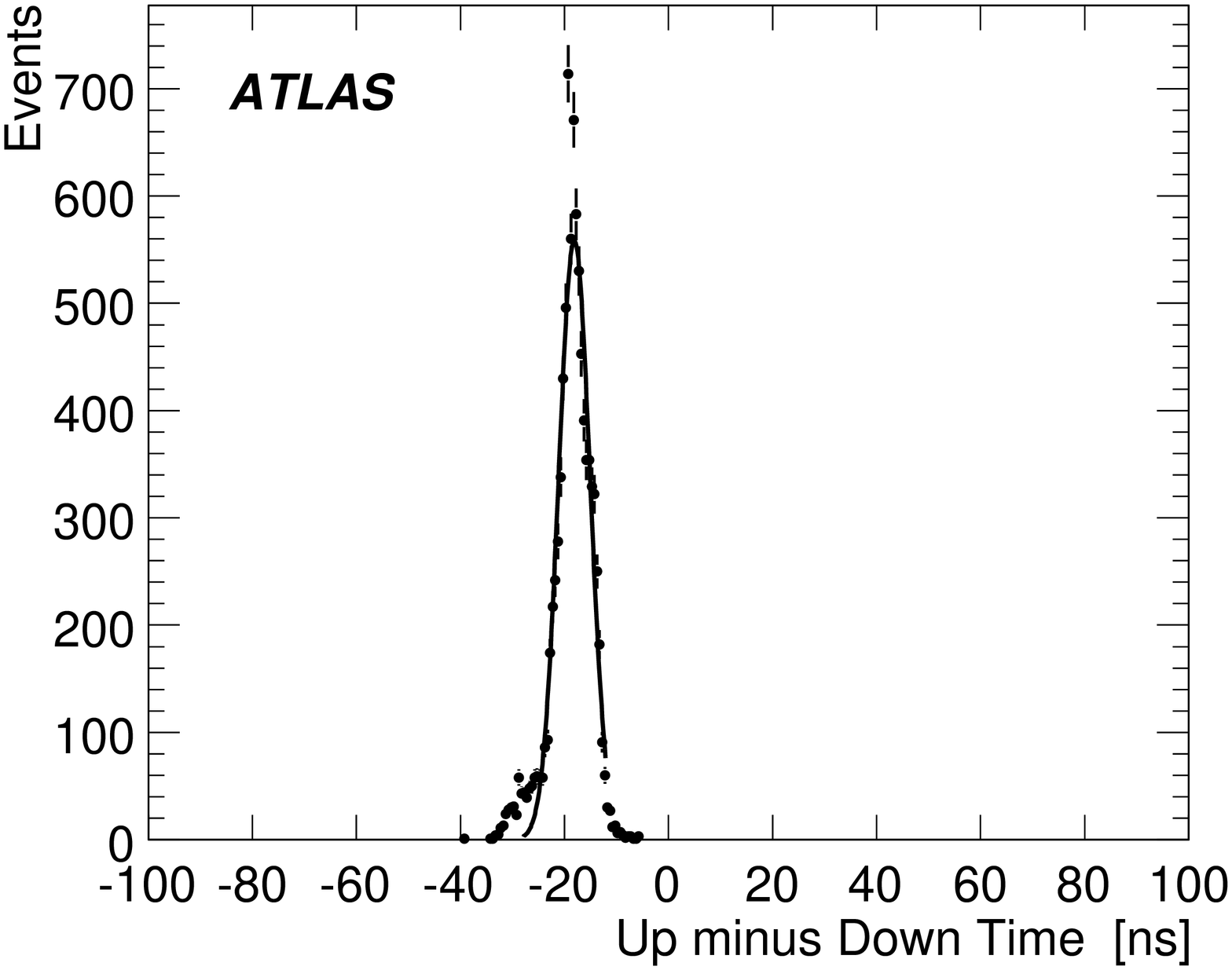}
&

  \includegraphics[height=.3\textheight,angle=270]{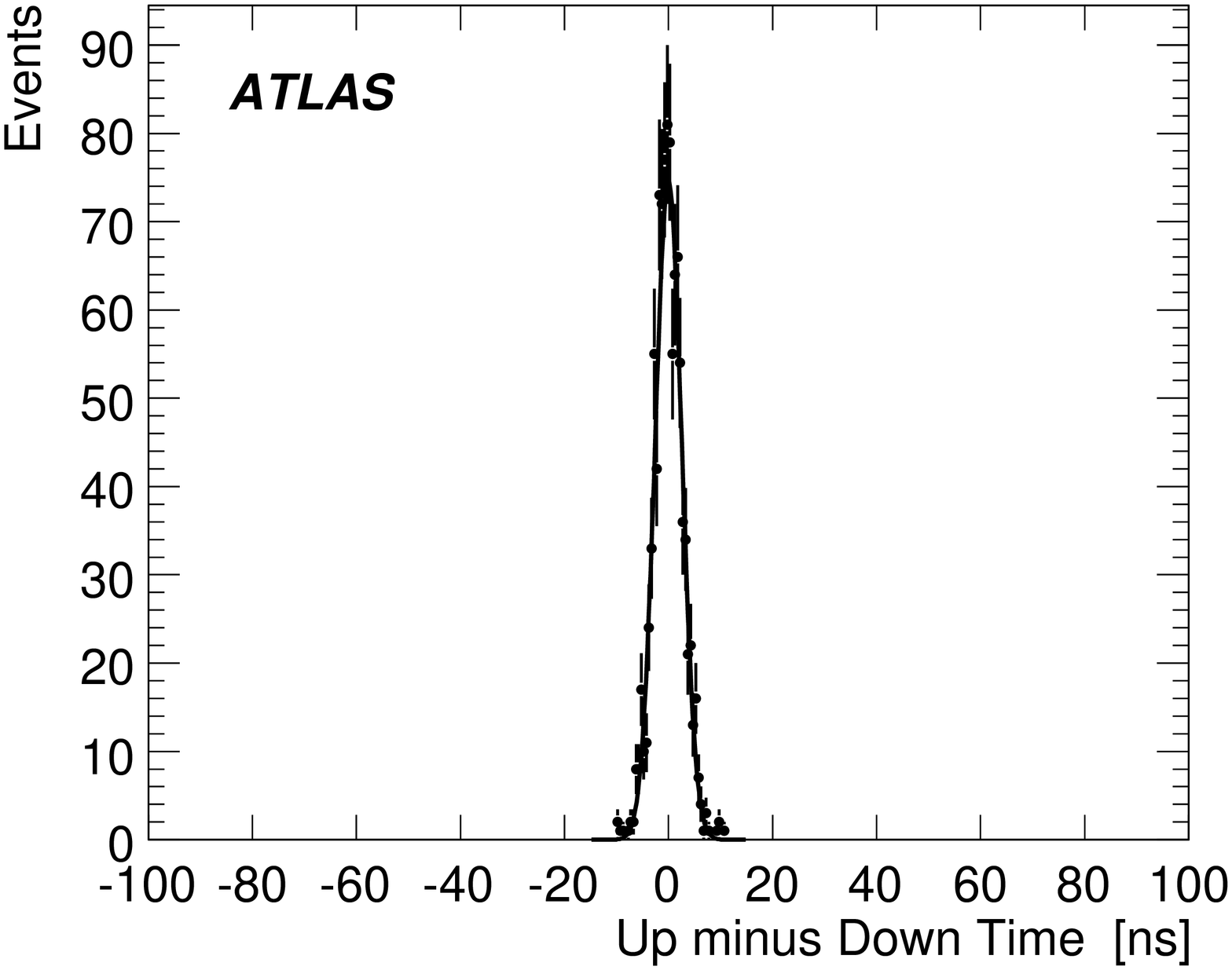}
\end{tabular}
  \caption{Time-of-Flight for a simulated Monte Carlo sample of cosmic ray muons (left) and QCD dijet events (right)}

\end{figure}

\section{Summary}

Cosmic ray data analysis is a crucial and important step for ATLAS commissioning. The techniques developed can be used in the first LHC data analysis. In particular, the use of calorimeter timing was explored. We first showed the importance of these studies for commissioning analysis in ATLAS. We crosschecked LASER run studies and helped improving PMT inter-module timing calibrations. It was also shown how such techniques can be used as tools for fake $\met{}$ rejection, an important result concerning searches for new physics relying on large $\met{}$, such as supersymmetry.


\begin{theacknowledgments}
I woulk like to thank Richard Teuscher and Gregor Herten for their suggestions for this talk. I would like to thank the Tile Calorimeter Community for the opportunity. I would also like to thank Jose Maneira and Cristophe Clement for their reading and suggestions for this document. This work was supported by the Federal Ministry of Education and Research (Germany).

\end{theacknowledgments}

\end{document}